\documentclass{article}
\usepackage{spconf,amsmath,graphicx, hyperref}

\usepackage{enumitem}
\setlist{nosep, leftmargin=14pt}

\usepackage{mwe} 

\title{Decoding Radiologists' Intentions: A Novel System for Accurate Region Identification in Chest X-ray Image Analysis}
\hypersetup{
    colorlinks=true,
    linkcolor=black,  
    urlcolor=blue,    
    citecolor=black   
}

\name{Akash Awasthi, Safwan Ahmad\textsuperscript{*},Bryant Le\textsuperscript{*}, Hien Nguyen\thanks{The preprocessed data is available  \href{https://github.com/a04101999/DECODING-RADIOLOGISTS-INTENTIONS-A-NOVEL-SYSTEM-FOR-ACCURATE-REGION-IDENTIFICATION-IN-CHEST-X-RAY/tree/main}{here}}}
\address{ University of Houston \\
            Electrical and Computer Engineering Department, \textsuperscript{*} Computer Science Department \\
            }
\begin{document}
%

\maketitle
\begin{abstract}
In the realm of chest X-ray (CXR) image analysis, radiologists meticulously examine various regions, documenting their observations in reports. The prevalence of errors in CXR diagnoses, particularly among inexperienced radiologists and hospital residents, underscores the importance of understanding radiologists' intentions and the corresponding regions of interest. This understanding is crucial for correcting mistakes by guiding radiologists to the accurate regions of interest, especially in the diagnosis of chest radiograph abnormalities. In response to this imperative, we propose a novel system designed to identify the primary intentions articulated by radiologists in their reports and the corresponding regions of interest in CXR images. This system seeks to elucidate the visual context underlying radiologists' textual findings, with the potential to rectify errors made by less experienced practitioners and direct them to precise regions of interest. Importantly, the proposed system can be instrumental in providing constructive feedback to inexperienced radiologists or junior residents in the hospital, bridging the gap in face-to-face communication. The system represents a valuable tool for enhancing diagnostic accuracy and fostering continuous learning within the medical community.
\end{abstract}
\section{Introduction}
\label{sec:intro}

Chest radiographs stand as one of the most commonly conducted imaging examinations globally\cite{ccalli2021deep}. Despite their prevalence, the interpretation of these images is susceptible to errors, with an estimated 4\% \ incidence of radiologist errors in typical cases encountered in practice\cite{bruno2015understanding}. Radiologists analyze CXR images by scrutinizing various regions, generating findings that are often recorded in reports. Unfortunately, errors in CXR image diagnoses are not uncommon, particularly among inexperienced radiologists or hospital residents. Such errors may arise from misinterpretation of the image or inaccuracies in diagnosing specific diseases.

In response, we propose a system designed to identify the primary intentions of radiologists in their reports and the corresponding regions of interest in CXR images, elucidating the visual meaning behind their textual findings. This system holds promise in rectifying mistakes made by inexperienced radiologists, guiding them to the correct regions of interest. Additionally, senior radiologists can utilize this system to direct residents or less experienced colleagues to the appropriate regions of interest for specific diseases, thereby saving time and enabling more efficient feedback.

Moreover, our proposed system has potential applications in the development of training programs for newly joined radiologists and can serve as a submodule for creating automated error correction systems. Comprising two submodules, TGID and RE, our system draws inspiration from the Dense Video Captioning (DVC) task\cite{tian2018audio}. Notably, our work represents the first attempt at applying Dense Video Captioning to medical data.

The key contributions of this paper include:

\begin{enumerate}

\item Development of a novel system for comprehending the intentions of radiologists alongside the corresponding regions of interest.
\item Introduction of evaluation strategies for assessing the proposed model.
\item Pioneering a new task known as radiologist intention detection within the medical domain.

\end{enumerate}

\section{Methodology}
\label{ssec:methodology}

\begin{figure}[ht]
  \centering
  \centerline{\includegraphics[width=\linewidth]{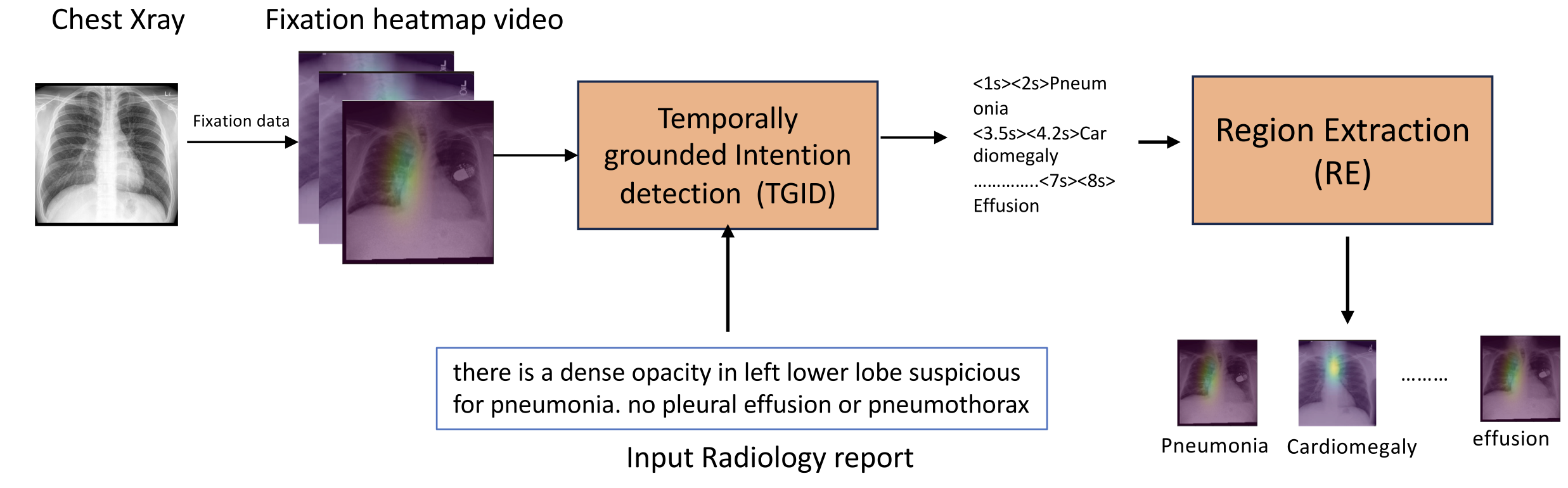}}
  \caption{Our proposed system to detect radiologists intention with corresponding ROI}
  \label{fig:fig1}
\end{figure}

Our proposed system comprises two primary modules: 1) Temporally Grounded Intention Detection (TGID) and 2) Region Extraction (RE). Illustrated in Fig \ref{fig:fig1} , the TGID module utilizes the fixation heatmap video and the time steps embedded in the radiology report as inputs. It then predicts the main intentions in the radiology report with the corresponding temporal grounding or time steps. Refer to Fig \ref{fig:fig2} for a detailed architecture of this module.

The Region Extraction (RE) module utilizes the predicted time steps (start and end times) and the identified intention to extract clips from the input video, containing multiple frames. Subsequently, we compute the mean of all images within the extracted clip to determine a representative image for the region of interest associated with the intended purpose. It's important to note that the RE module is a straightforward search algorithm reliant on TGID predictions and is not the primary focus of our contribution.
\vspace{-0.5cm}
\subsection{Architecture}
\label{ssec:architecture}

\begin{figure}[ht]
\begin{minipage}[t]{\linewidth}
  \centering
  \centerline{\includegraphics[width=\linewidth, height = 2in]{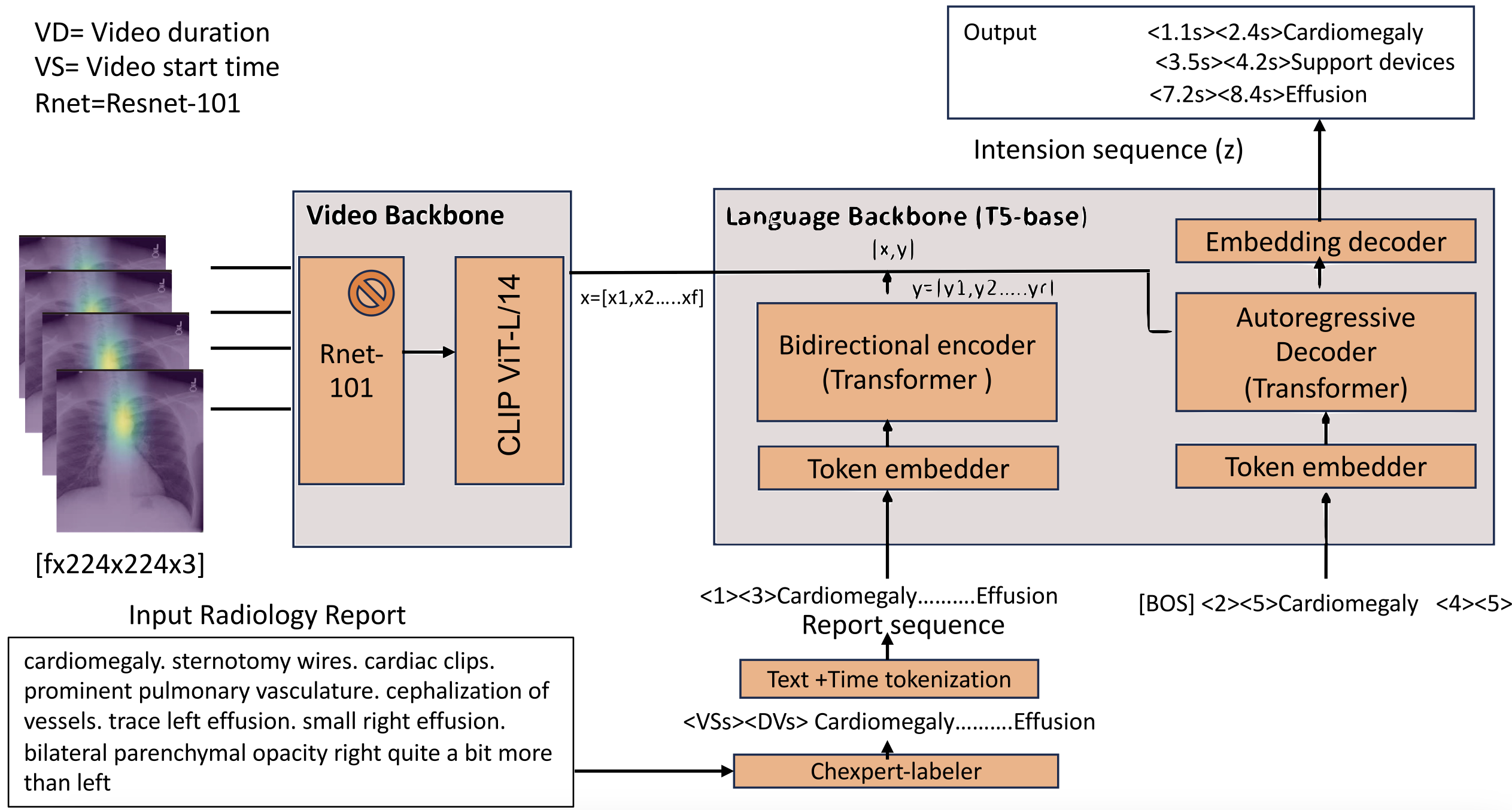}}
  \caption{TGID module overview: A sequence to sequence model which takes video features  and summarized radiology report with appended time tokens as input and output the intention sequence with temporal grounding }
  \label{fig:fig2}
\end{minipage}
\end{figure}

The TGID module serves as the core of our system by forecasting temporally grounded intentions. The intricate design of this module is depicted in the Fig \ref{fig:fig2} . As illustrated, the proposed architecture comprises two integral components: the Video Backbone and the Language Backbone. We employed Chexpert-labeler\cite{irvin2019chexpert} to condense the radiology report into main 14 chexpert labels\cite{irvin2019chexpert}. To enhance the summary, we added the start and end times, with the end time representing the video duration and the start time set at 1.1 seconds. This choice is grounded in our observation that radiologists typically commence speaking after 1.1 seconds upon viewing the video, a value derived from the analysis of the EGD-CXR and REFLACX datasets.

\subsection{Video Backbone}
\label{ssec:videobackbone}

The Video Backbone plays a pivotal role in extracting features from the input video. It comprises of a spatial encoder followed by a temporal encoder, operating on a sequence of 'f' frames. Utilizing a pretrained Resnet-101\cite{ccalli2021deep} as the spatial encoder, we extract individual frame features, considering the spatial characteristics of each frame in the video. The input set consists of videos with dimensions 'f × h × w × c,' where 'h,' 'w,' and 'c' represent the height, width, and number of channels of each frame. The spatial encoder processes each frame independently, and we maintain the spatial backbone as frozen to minimize computational costs and parameter count in the overall model.

The spatial encoder generates a two-dimensional array, with the first dimension representing the number of frames and the second representing the embedding dimension. Although each video may have a varying number of frames, we limit our consideration to the features of the first 100 frames. To accommodate videos with fewer than 100 frames, we pad the feature extraction output from Resnet with zeros.

For the temporal encoder, we employ a pretrained CLIP ViT-L/14\cite{dosovitskiy2020image, radford2021learning} transformer to produce contextualized embeddings, contributing to the comprehensive feature representation of the input video.
\subsection{Language Backbone}
\label{ssec:languagebackbone}

Our language backbone\cite{yang2023vid2seq} is built on the T5\cite{radford2021learning} , employing an encoder-decoder architecture. We initialized both the text encoder and decoder with the t5-base model, which underwent pretraining on web text corpora with a denoising loss.

\subsubsection{Text and  Time tokenization}
\label{ssec:textencoder}

We employ the SentencePiece tokenizer\cite{kudo2018sentencepiece} with a vocabulary size of V = 32,128. Beginning with text tokenization, we enhance it by appending two additional time tokens. The time tokenization process follows the equation outlined below.

\subsubsection{Text Encoder}
\label{ssec:textencoder1}

It accepts a report sequence as input, where the report sequence comprises 'r' tokens denoted as 'y' belonging to the set \( y \in \{1, \ldots, v + n\}^r \) Here, 'v' represents the vocabulary size of text, 'n' is the size of time tokens, and 'r' stands for the total number of tokens in the report sequence. The text encoder includes an embedding layer responsible for independently embedding each token, producing a semantic embedding of size 'rxd'. Subsequently, a transformer encoder calculates contextualized embeddings of size 'rxd', with 'd' representing the hidden dimension.

\subsubsection{Text Decoder}
\label{ssec:textdecoder}

Comprising a transformer decoder and an embedding layer, the system generates an intention sequence with associated temporal grounding, referred to as the intention sequence. The transformer decoder, operating in a causal manner, engages in cross-attention with the encoder output and all previously generated tokens. Additionally, it conducts self-attention across the entire set of previously generated tokens. An embedding decoder is applied atop the transformer text decoder, predicting the probability distribution over the joint vocabulary of text and time tokens. This enables the model to anticipate the next token in the report sequence.

\subsection{Pretraining \& Finetuning}
\label{ssec:training}

For the pretraining of TGID, we employed the ActivityNet Captions dataset\cite{wang2018bidirectional}, which features approximately 20,000 untrimmed videos capturing diverse human activities. These videos include temporally grounded events accompanied by transcribed speech sentences and timestamps. Due to a scarcity of fixation videos and corresponding transcriptions in medical domain, we turned to these publicly available videos with speech transcriptions for our pretraining efforts. This initial training of the TGID block is instrumental in enabling the model to comprehend long-term relationships among various speech segments. Our approach involved incorporating two training objectives for this model: a generative objective and a denoising objective, both detailed in this work\cite{yang2023vid2seq}.

Subsequently, during the finetuning stage, the model is refined to predict the intention sequence by considering both the radiology report sequence and the visual sequence. The fine-tuning objective is derived from the maximum likelihood objective, elaborated upon in this context\cite{yang2023vid2seq}.


\section{Datasets \& Expertimentation}
\label{sec:related}
For this study, we utilized the EGD-CXR\cite{karargyris2021creation}  and REFLACX\cite{bigolin2022reflacx} datasets. The EGD-CXR dataset comprises 1,083 chest X-ray images reviewed by a radiologist using an eye-tracking system. Meanwhile, the REFLACX dataset encompasses 3,032 cases with synchronized eye-tracking and transcription pairs, annotated by five radiologists. Our approach involved generating fixation heatmaps overlaid on X-ray images by leveraging eye gaze data, providing a dynamic representation of the gaze movement over the CXR image.

The speech transcription data from EGD-CXR and REFLACX was employed to create the intention sequence file, incorporating temporal grounding. This served as the ground truth during the model's finetuning on fixation heatmap videos and radiology reports.

Our pretraining phase involved 1,000 epochs with a batch size of 8, consuming a day on 8 Tesla GPUs. For finetuning, we amalgamated the REFLACX and EGD-CXR datasets, yielding 4,115 samples. Among these, 2,115 were allocated for finetuning, 1,000 for validation, and 1,000 for testing. The model was trained on this dataset, consisting of fixation heatmap videos, radiology reports, and temporally grounded intention sequences as ground truth. We employed Adam as the optimizer, with a batch size of 2 for both validation and training.

Notably, despite planning for 600 epochs of finetuning, the model demonstrated effective learning within approximately 50 epochs.
\begin{figure*}[ht]
  \centering
  \centerline{\includegraphics[width=0.95\linewidth]{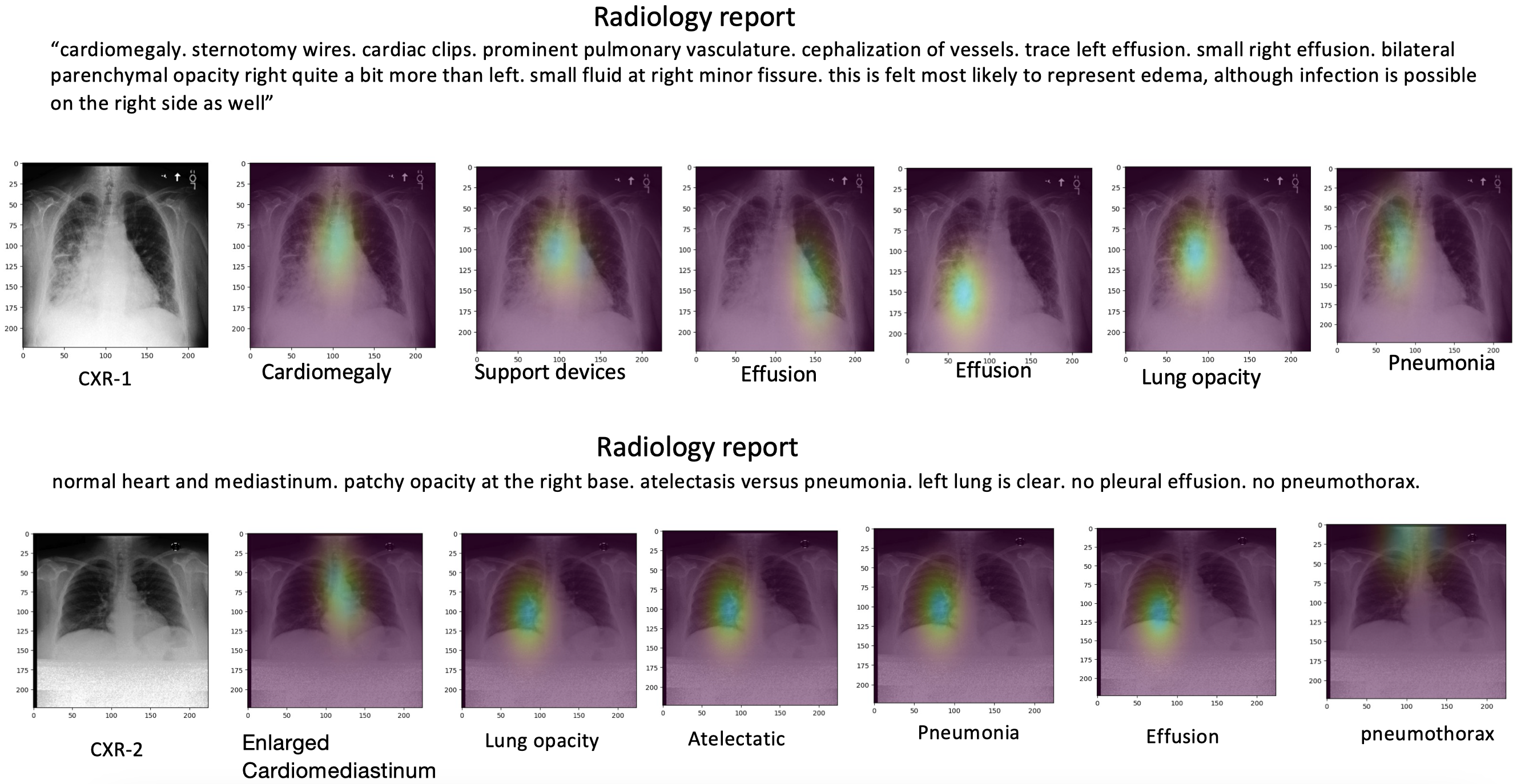}}
  \caption{Anticipated intentions of radiologists and their corresponding regions of interest are illustrated. This figure depicts the specific areas within the image that radiologists focus on for each diagnosis }
  \label{fig:fig3}
\end{figure*}
\section{Results \& Discussions}
\label{sec:results}
We employ natural language generation metrics to assess the proposed model's text generation capability, comparing it with state-of-the-art methods. Additionally, we evaluate its temporal groundings by analyzing time delay errors.

\subsection{State of the Art Comparison}
\label{ssec:comparison}

The model in question predicts text containing various intentions along with their corresponding time steps. Natural Language Generation (NLG) metrics serve as a valuable tool in gauging the model's text generation proficiency. While NLG metrics may not be the optimal measure for evaluating predicted time steps for each intention, they provide a holistic indication that the model is generating meaningful output. The accompanying table displays n-gram Blue scores and CIDER scores for various state-of-the-art methods, indicating that our model excels in generating intentions with accurate temporal grounding. 

\begin{table}[h]
\normalsize
\resizebox{\linewidth}{!}{
\begin{tabular}{lcccccc} 
\hline
Method         & Backbone          & Blue-1 & Blue-2 & Blue-3 & Blue-4 & CIDER \\ \hline
PDVC \cite{wang2021endtoend}     & V (CLIP)           & 0.21   & 0.17   & 0.13   & 0.11   & 1.8   \\
VTransformer \cite{zhou2018endtoend}  & V (ResNet-200) + F & 0.11   & 0.09   & 0.07   & 0.05   & 0.08      \\
MFT \cite{xiong2018move}   & V + F (TSN)       & 0.09    & 0.07   & 0.04   & 0.03   & 0.4      \\
MART \cite{lei2020mart}   & V (ResNet-200) + F     & 0.19    & 0.15    & 0.12   & 0.10   & 1.1    \\
Transformer-XL \cite{dai2019transformerxl} & V (ResNet-200) + F & 0.17   & 0.14   & 0.10   & 0.08   & 1.0   \\
Our    & ResNet-101          & 0.47   & 0.46   & 0.45   & 0.43   & 3.6   \\ \hline
\end{tabular}
}
\caption{Comparison to the SoTA for Intention temporal grounding. V/F/O refers to visual/flow/object features
 }
\end{table}

\begin{figure}[htb]
\begin{minipage}[b]{\linewidth}
  \centering
  \centerline{\includegraphics[width=\linewidth]{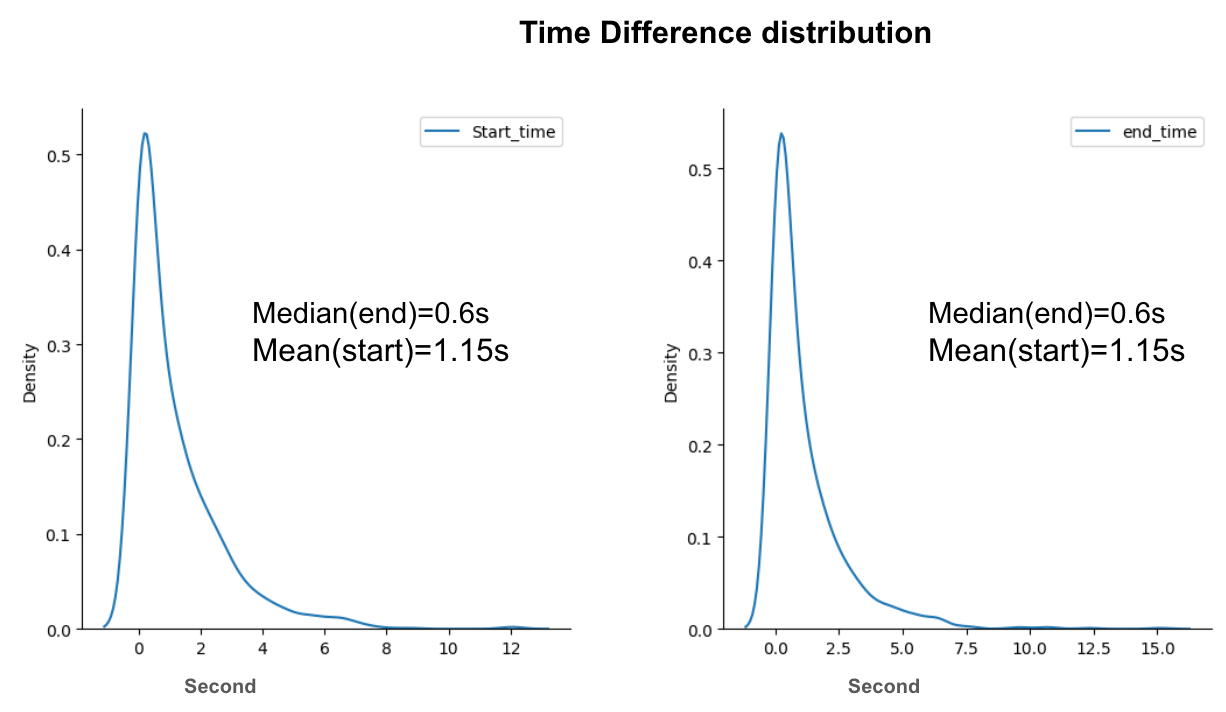}}
  \caption{Distribution plot of  differnce between the predicted and true start and end time for each intention in the test set }
  \label{fig:fig4}
\end{minipage}
\end{figure}

We computed the disparity between predicted and actual start times for each intended prediction in the test dataset, extending the same analysis to end times. In the subsequent presentation, we depict the distribution of these differences in two distinct plots, revealing a centralization around 0.6 sec for both start and end times. As shown in figure \ref{fig:fig4} This indicates that the proposed system adeptly predicts temporal grounding for each intention.

\subsection{Time Delay Error for Each Disease}
\label{ssec:timedelay}

The primary aim of this proposed method is to discern the temporal groundings of intentions. Consequently, the initial assessment revolves around evaluating the accuracy of predicted time steps. To grasp the disparity between predicted and true time steps, we introduce a time delay error metric. Given the skewed distribution, we opt for the median as the estimator, defining the Median Time Delay Error (MTDE) with the provided mathematical formula.

\begin{equation}
\text{MTDE} = \text{median}(\text{gt(start)} - \text{pred(start)})
\end{equation}

As presented in Table 2, we showcase the median time delay error. Notably, most diseases exhibit minimal time delay errors, signifying the model's proficiency in accurately detecting diseases with the appropriate temporal grounding in the video. Some diseases, however, show higher errors, potentially stemming from misalignment between Automatic Speech Recognition (ASR) and video data during training. Further investigation reveals that diseases with elevated MTDEs are often those where diagnosis by a physician takes longer compared to those with lower MTDEs. It is noteworthy that the proposed method may face challenges in generating larger intervals. This MTDE metric serves as an evaluation of the model's effectiveness in identifying the region of interest for the predicted intention or disease.

\begin{table}[h]
    \centering
    \resizebox{\linewidth}{!}{
    \begin{tabular}{lccc}
        \hline
        Disease (Intension) & (MTDE) (sec) & Disease (Intension) & (MTDE) (sec) \\
        \hline
        Pneumonia           & 0.88         & Atelectasis         & 1.7          \\
        Effusion            & 0.88         & Lung Opacity        & 0.5          \\
        Pneumothorax        & 0.87         & Lung Lesion         & 0.7          \\
        Edema               & 1.8          & Normal Heart        & 0            \\
        Cardiomegaly         & 0          & Support Devices        & 0            \\
        \hline
    \end{tabular}
    }
    \caption{MTDE for each label predicted by the our system }
    \label{tab:my_label}
\end{table}
\vspace{-0.5cm}
\subsection{Visualization of the intentions}
\label{ssec:visualization}

Presented here are visualizations of predicted intentions along with their corresponding regions in images, indicative of what the radiologist focuses on during X-ray image diagnosis. In Figure \ref{fig:fig3}, the complete report is displayed in the first row, followed by the diagnosis label and the associated highlighted regions generated by the proposed system. It's important to note that this visualization pertains to two randomly selected X-ray images from the test set. 
\section{Acknowledgement} This work was supported in part by the National Institutes of Health under Grant 1R01CA277739. The content is solely the responsibility of the authors and does not necessarily represent the official views of the National Institutes of Health.

\bibliographystyle{IEEEbib}
\bibliography{strings,refs}

\end{document}